\documentclass[epj]{svjour}
\usepackage{graphics}
\begin{document}
\title{Continuous Percolation Phase Transitions of Two-dimensional Lattice Networks under a Generalized Achlioptas Process}
%\subtitle{Do you have a subtitle?\\ If so, write it here}

\author{Maoxin Liu\inst{1} \and Jingfang Fan\inst{1}\and Liangsheng Li\inst{2}\and Xiaosong Chen\inst{1}% etc
% \thanks is optional - remove next line if not needed
\thanks{\emph{e-mail:chenxs@itp.ac.cn}}%
}                     % Do not remove
%
%\offprints{}          % Insert a name or remove this line
%
\institute{Institute of Theoretical Physics, Chinese Academy of Sciences,
P.O. Box 2735, Beijing 100190, China\and Key Laboratory of Cluster
Science of Ministry of Education and Department of Physics,
Beijing Institute of Technology, Beijing 100081, China}
\date{Received: date / Revised version: date}
% The correct dates will be entered by Springer
%

\abstract{The percolation phase transitions of two-dimensional lattice networks under a generalized Achlioptas
process (GAP) are investigated. During the GAP, two edges are chosen randomly from the lattice and the edge
with minimum product of the two connecting cluster sizes is taken as the next occupied bond with a probability $p$.
At $p=0.5$, the GAP becomes the random growth model and leads to the minority product rule at $p=1$. Using the finite-size scaling analysis, we find that the percolation phase transitions
of these systems with $0.5 \le p \le 1$ are always continuous and their critical exponents depend on $p$.
Therefore, the universality class of the critical phenomena in two-dimensional lattice networks under the GAP
is related to the probability parameter $p$ in addition.
 % end of PACS codes
} %end of abstract
\maketitle

\section{Introduction}
\label{Introduction}

The percolation phase transition concerns the formation of a
macroscopic component in systems on both lattices and networks
\cite{StaufferBook}. It provides a model for the onset of a
macroscopic component in random media \cite{StaufferBook} and social
networks \cite{Solomon}. It was widely believed that the
percolation transition is a typical continuous phase transition
for various networks \cite{Dorogovtsev}. However, Achlioptas,
D'Souza, and Spencer \cite{Achlioptas} found recently that the
percolation phase transition in random network becomes
discontinuous (first-order) under the Achlioptas process (AP),
where the edge with minimum product of cluster masses is connected
from the two randomly chosen unoccupied edges. The Achlioptas
process suppresses the appearance of larger cluster and
discourages the formation of a giant component, which has the size
comparable with the number of vertices $N$. The percolation
phase transition is delayed by this Achlioptas process and becomes sharper.
It was argued by them\cite{Achlioptas} that the phase transition is discontinuous and named as
an explosive percolation. Later, the Achlioptas process was introduced to the two-dimensional
regular lattice networks \cite{ZiffPRL09} and scale-free networks
\cite{ChoPRL09,RadicchiPRL09}.  It was claimed that the explosive
percolation was found both in lattice \cite{ZiffPRL09} and in scale-free
networks\cite{ChoPRL09,RadicchiPRL09}.

In the article of Achlioptas et al.\cite{Achlioptas}, the step interval $\Delta$ between the size of the largest component $S_1=N^{1/2}$ and $S_1=0.5N$ is used as the criterion of continuous or discontinuous phase transition. Later, Ziff applied this criterion to the two-dimensional regular lattice networks\cite{ZiffPRL09,ZiffPRE10}. It was found that the size-dependence of $\Delta$ in lattice network is quite different from that in random network. It is not well established that the first-order phase transition can be distinguished from the continuous phase transition by the size-dependence of $\Delta$.
It was argued by da Costa et al.\cite{CostaPRL10} that the explosive percolation transition, under their modified Achlioptas process, is actually continuous. Recently, Riordan et al.\cite{RiordanArXiv} show mathematically that all Achlioptas process have continuous phase transitions. The finite-size behavior of the order-parameter distribution function has been used as the evidence of both discontinuous\cite{TianArXiv} and continuous\cite{GrassbergerArXiv} phase transition. It is also argued with the finite-size scaling that the explosive percolation is  continuous \cite{RadicchiPRE10,FortunatoArXiv}. This controversy about the character of explosive percolation is going on and calls for further investigations.

In this paper, we investigate the percolation phase transition in two-dimensional lattice network under a generalized Achlioptas process (GAP), which will be introduced in the next section. The generalized Achlioptas process is characterized by a probability parameter $p$.  The GAP becomes the random growth model at $p=1/2$ and the minority product rule at $p=1$. Using the finite-size scaling analysis, our Monte Carlo simulation results demonstrate clearly that the percolation phase transition in two-dimensional lattice networks under the GAP is continuous. It will be shown that the critical exponents and therefore the universality class of the continuous percolation phase transition depend on the probability parameter $p$.

Our paper is organized as follows. In the next section, we introduce a
generalized Achlioptas process in two-dimensional lattice network. In Section 3, we investigate the critical points of two-dimensional lattice network under the GAP and their critical exponents with the use of finite-size scaling. In Section 4, the finite-size scaling function of the ratio $S_2/S_1$ is obtained at different probability parameter $p$, where $S_2$ and $S_1$ are the size of the second largest and the largest cluster in the network. The universality class of the critical points in our model is discussed in Section 5. Finally we make some conclusions in Section 6.

\section{Two-dimensional lattice network under the GAP }
\label{Model}
We consider a two-dimensional square lattice with
size $L\times L$ and periodic boundary conditions in both
directions.  There are $N=L^2$ vertices in this lattice. We
introduce a generalized Achlioptas process for adding edges into
this lattice. In the generalized Achlioptas process, two edges are picked up
randomly at each step. Each edge is connected with two clusters. The edge with the minimum
product of the cluster sizes is chosen and added into the lattice with a probability $p$,
where $0\le p \le 1$. Correspondingly, another edge is chosen with a probability $1-p$.
At $p=0.5$, the GAP is equivalent to the classic Erd\"os-R\'enyi (ER)
rule, where edges are picked up randomly. The two-dimensional
square lattice with the ER rule is actually the two-dimensional bond percolation (BP) model.
The GAP at $p=1$ is the product rule (PR) of Ref.\cite{Achlioptas} and our model becomes the PR model on the two-dimensional regular lattice.

In our Monte Carlo simulations, there are $N$ isolated
vertices in a two-dimensional lattice at the beginning and then edges are
added into the lattice through the GAP. With the edges added, we obtain a network
in the lattice. The lattice network can be characterized by a reduced edge number
$r\equiv N_r/N$, where $N_r$ is the number of the edges added.

We have used the algorithm of Newmann and Ziff
\cite{newmann1,newmann2} in our Monte Carlo simulations. For the investigations related only to the largest cluster of lattice networks in Figs. 3, 5 and 7, the linear sizes $L=32,~64,~128,
~256,~512$, and $1024$ are taken. When the second largest cluster in the lattice is taken into account in addition, only three linear sizes $L=64,~128$, and $256$ are taken in Figs. 2, 4 and 6. To get enough samples for the average of each simulation, different steps are taken for the simulation of different system size. In our Monte carlo simulations, we run $10,000,000$ steps for $L=32$ until to $6,400,000$ steps for $L=1024$.

\begin{figure}
% Use the relevant command for your figure-insertion program
% to insert the figure file.
% For example, with the option graphics use
\resizebox{0.5\textwidth}{!}{%
  \includegraphics{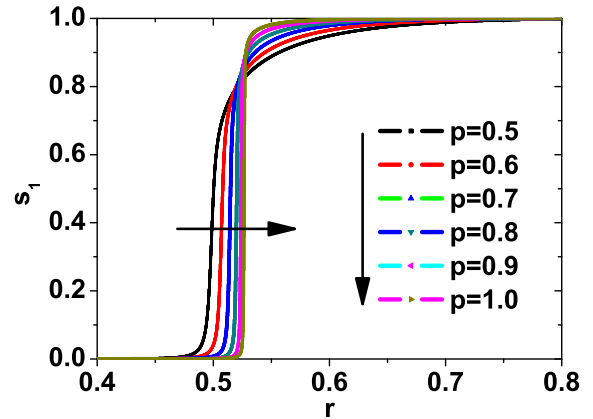}
}
% If not, use
%\vspace{5cm}       % Give the correct figure height in cm
\caption{Reduced size of the largest cluster, defined in Eq.
(\ref{ratio1}), as a function of the reduced edge number $r$ at the lattice size
$L=1024$ and probability parameters $p=0.5,~0.6,~0.7,~0.8,~0.9$, and $1.0$. }
\label{s}       % Give a unique label
\end{figure}

For the cluster ranked $R$ and with size $S_R (r,L;p)$, we defined its reduced size as
\begin{equation}
\label{ratio1} s_R(r,L;p)\equiv S_R (r,L;p)/N.
\end{equation}
In Fig.\ref{s}, we shown the reduced size $s_1 (r,L;p)$ of the largest cluster.

In the Monte Carlo simulations of these data, we take the lattice size $L=1024$ and the
probability parameter $p=0.5,~0.6,~0.7,~0.8,~0.9$, and $1.0$. At small $r$, the
reduced size of the largest cluster is nearly zero. When  $r$ is large enough, the reduced size $s_1$
becomes finite. This indicates the formation of
a macroscopic component. Therefore, there is a percolation phase transition in the lattice network.
The transition value of the reduced edge number $r_c$ depends on the probability parameter $p$.
It is shown in Fig.\ref{s} that $r_c$ increases with the probability parameter $p$. This is plausible since a larger $p$ means stronger suppression of larger cluster and therefore the later appearance of a macroscopic component. In the following, we will try to verify that these percolation phase transitions are continuous or not.

\section{Critical points of two-dimensional lattice networks under GAP}
\label{critical point}

If the percolation phase transitions above
were continuous, the reduced size of the cluster ranked $R$ should
follow the finite-size scaling form \cite{PF1984,privman}
\begin{equation}\label{s-scale}
s_R(r,L;p)=L^{-\beta/\nu}\tilde{s}_R(tL^{1/\nu};p),
\end{equation}
where $t=(r-r_c)/r_c$ characterizes the deviation from the
critical point $r_c$ and $\nu$ is the critical exponent of the correlation length
$\xi=\xi_0 |t|^{-\nu}$. This finite-size scaling form is supposed to be valid in the asymptotic critical region where $L\gg$ lattice spacing and $|t|\ll 1$. Outside the asymptotic region, additional correction terms should be taken into account.

For the largest cluster of lattice network, we have the finite-size
scaling form
\begin{equation}\label{eq:S1}
s_1(r,L;p)=L^{-\beta/\nu}\tilde{s}_1(tL^{1/\nu};p).
\end{equation}
In the bulk limit $L \to \infty$, the reduced size of the largest
cluster becomes
\begin{equation}
s_1(r,\infty;p)=0
\end{equation}
for $r < r_c$ and
\begin{equation}
s_1(r,\infty;p)=a_p\; t^{\beta}
\end{equation}
for $r > r_c$. The emergent macroscopic
component is characterized by the critical exponent $\beta$. The smaller the critical exponent $\beta$ is, the larger is the macroscopic component.

Near the critical point, the reduced size of the second largest cluster can
be written also in a finite-size scaling form
\begin{equation}\label{eq:S2}
s_2(r,L;p)=L^{-\beta/\nu}\tilde{s}_2(tL^{1/\nu};p).
\end{equation}
Using Eqs. (\ref{ratio1}, (\ref{eq:S1}), and (\ref{eq:S2}), we can obtain the finite-size scaling form of the ratio
\begin{equation}\label{eq:S2/S1}
S_2/S_1=\tilde{s}_2(tL^{1/\nu};p)/\tilde{s}_1(tL^{1/\nu};p)\equiv U(tL^{1/\nu};p).
\end{equation}
At the critical point $t=0$, the ratio
\begin{equation}\label{ratioc}
\left.S_2/S_1\right|_{r=r_c}=U(0;p),
\end{equation}
which is independent of the system size $L$. Therefore, the curves
of $S_2/S_1$ at different system size $L$ have a cross-point at $r=r_c$.
The critical point corresponds the fixed point of $S_2/S_1$, which can be
used to determine the critical point of our system.

The logarithm of Eq. (\ref{eq:S1}) can be expressed as
\begin{equation}\label{eq:lnS}
\ln s_1(r,L;p)=-(\beta/\nu) \ln L+\ln \tilde{s}_1(tL^{1/\nu};p).
\end{equation}
At the critical point $r =r_c$, we have
\begin{equation}\label{eq:lnS1}
\ln s_1(r_c,L;p)=-(\beta/\nu) \ln L+\ln \tilde{s}_1(0;p),
\end{equation}
which is a straight line with respect to $\ln L$. We can use this property to determine the critical point $r_c$ of the lattice networks also. From the slope of this straight line, the critical exponent ratio $\beta/\nu$ can be determined.

In the following, we use both the fixed point of $S_2/S_1$ and the linear
dependence of $\ln s_1$ on $\ln L$ as the criterion to determine the critical point of the two-dimensional lattice networks under the generalized
Achlioptas process. If we have reached the asymptotic critical region, both $s_1$ and $s_2$ satisfy the finite-size scaling form in Eqs.(\ref{eq:S1}) and (\ref{eq:S2}). The critical reduced edge numbers $r_c$ obtained from $S_2/S_1$ and $\ln s_1$ should be equal. On the other hand, the consistence of $r_c$ obtained from two different methods can be used as the indicator of the accuracy of our simulation results.

\begin{figure}
% Use the relevant command for your figure-insertion program
% to insert the figure file.
% For example, with the option graphics use
\resizebox{0.5\textwidth}{!}{%
  \includegraphics{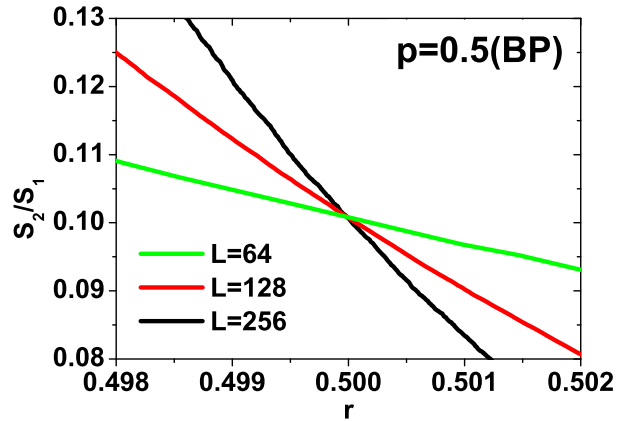}
}
% If not, use
%\vspace{5cm}       % Give the correct figure height in cm
\caption{Size ratio $S_2/S_1$ of the second largest to the largest cluster at $p=0.5$ and different $L$. There is a fixed point
at $r_c=0.5000\pm0.0002$.}
\label{ratiobp}       % Give a unique label
\end{figure}

For the generalized Achlioptas process with probability parameter $p=0.5$,
edges are added randomly into a two-dimensional lattice and our model becomes
the so-called bond percolation model. It is well known that the bond percolation model
has a continuous phase transition. We can determine the critical point
of this model by the two methods described above. In Fig.\ref{ratiobp}, the ratio $S_2/S_1$ is
shown as a function of the reduced edge number $r$ at different
system size $L$. A fixed point between $r=0.4998$ and $r=0.5002$ is found and is in full agreement with
$r_c=1/2$ of the bond percolation model. In Table \ref{tab1}, we denote the
critical point obtained from $S_2/S_1$ as $r_c^{(1)}=0.5000 \pm 0.0002$.

\begin{figure}
% Use the relevant command for your figure-insertion program
% to insert the figure file.
% For example, with the option graphics use
\resizebox{0.5\textwidth}{!}{%
\includegraphics{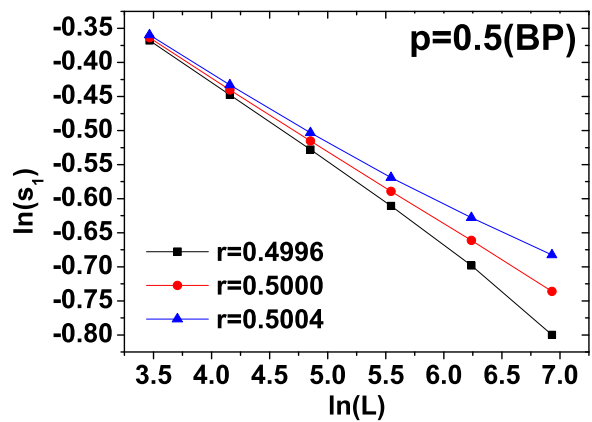}
}
% If not, use
%\vspace{5cm}       % Give the correct figure height in cm
\caption{Log-log plot of the reduced size $s_1$ of the largest cluster at $p=0.5$ and different $r$. We have chosen linear sizes $L=32,~64,~128,~256,~512$,
and $1024$.}
\label{slopebp}       % Give a unique label
\end{figure}

As we have discussed above, a critical point can be determined alternatively by the linear
relationship between $\ln s_1 $ and $\ln L$. In Fig.\ref{slopebp}, $\ln s_1 (r,L;p)$ at different reduced edge numbers are shown. At $r=0.4996$ and $r=0.5004$, the curves of $\ln s_1 (r,L;p)$ are curved and their curvatures have different sign. At $r=0.5$, the curve of $\ln s_1 (r,L;p)$ becomes a straight line. Therefore, we obtain the critical point $r_c^{(2)}=0.5000 \pm 0.0004$, which is in agreement with $r_c^{(1)}$. From the slope of
$\ln s_1 (r,L;p)$ at $r_c$, we get the critical exponent ratio $\beta/\nu=0.108$. Our Monte
Carlo simulation results agree very well with the exact results $r_c=1/2$ and $\beta/\nu=5/48$ of two-dimensional bond percolation model, for a review, see Ref. \cite{essam}.

\begin{figure}
% Use the relevant command for your figure-insertion program
% to insert the figure file.
% For example, with the option graphics use
\resizebox{0.5\textwidth}{!}{%
  \includegraphics{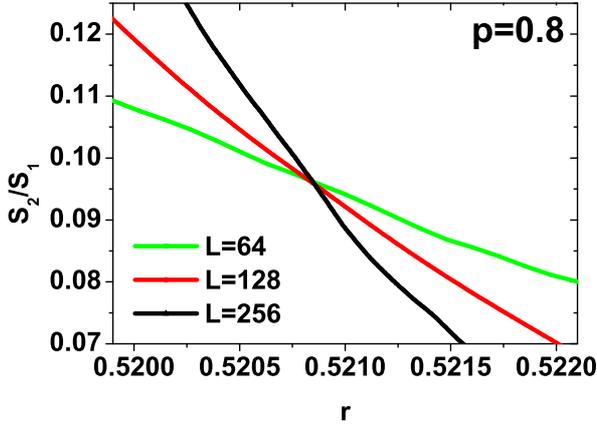}
}
% If not, use
%\vspace{5cm}       % Give the correct figure height in cm
\caption{Ratio $S_2/S_1$ at  $p=0.8$ and
different $L$. There is a fixed point at $r=0.5208 \pm 0.0001$.}
\label{ratio08}       % Give a unique label
\end{figure}

\begin{figure}
% Use the relevant command for your figure-insertion program
% to insert the figure file.
% For example, with the option graphics use
\resizebox{0.5\textwidth}{!}{%
\includegraphics{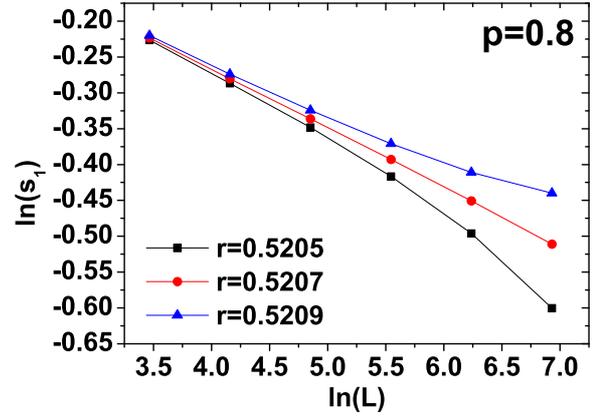}
}
% If not, use
%\vspace{5cm}       % Give the correct figure height in cm
\caption{Log-log plot of the reduced size $s_1$ of the largest cluster at $p=0.8$
and different $r$. We have chosen linear sizes $L=32,~64,~128,~256,~512$,
and $1024$.}
\label{slope08}       % Give a unique label
\end{figure}

For probability parameter $p=0.8$, the edge that minimizes the product of two connecting cluster sizes is added into the lattice with a probability $0.8$ from two randomly chosen edges. The connection of smaller clusters is favored. The critical point of this system is investigated by the fixed point of  $S_2/S_1$
and the linear dependence of $\ln s_1$ on $\ln L$. In Fig.\ref{ratio08}, the ratio $S_2/S_1$ is plotted
as a function of the reduced edge number $r$ at different system sizes $L$.  A fixed point of
$S_2/S_1$ is found. It is between $r=0.5207$ and $r=0.5209$. So there is
a continuous phase transition in this system and the critical point is at $r_c^{(1)}=0.5208 \pm 0.0001$.
Alternatively, this critical point can be determined from $\ln s_1 (r,L;p)$. In Fig.\ref{slope08}, $\ln s_1 (r,L;p)$ at $r=0.5205, 0.5207$, and $0.5209$ are shown. The curvature of $\ln s_1 (r,L;p)$ is negative at $r=0.5205$ and positive at $r=0.5209$. The function $\ln s_1 (r,L;p)$ at $r=0.5207$ can be described quite well by a straight line with zero curvature. So the critical point $r_c^{(2)}=0.5207 \pm 0.0002$, in agreement with $r_c^{(1)}$. The slope of the straight line at $r_c=0.5207$ gives the critical exponent ratio $\beta/\nu=0.081$.

%---------------------------------------------------------------

%---------------------------------------------------------------

\begin{figure}
% Use the relevant command for your figure-insertion program
% to insert the figure file.
% For example, with the option graphics use
\resizebox{0.5\textwidth}{!}{%
  \includegraphics{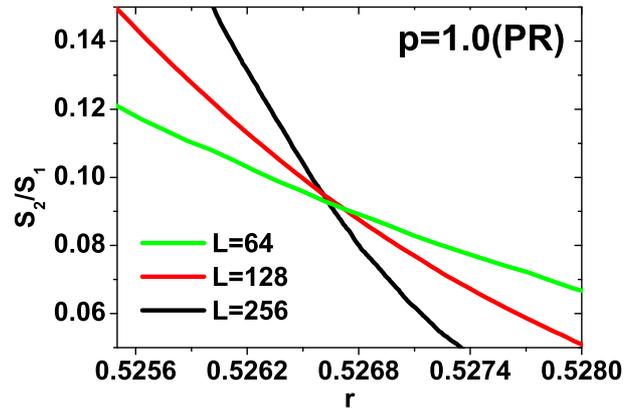}
}
% If not, use
%\vspace{5cm}       % Give the correct figure height in cm
\caption{Ratio $S_2/S_1$ at $p=1.0$ and different $L$ .
There is a fixed point at $r=0.5266 \pm 0.0001$.}
\label{ratiopr}       % Give a unique label
\end{figure}

\begin{figure}
% Use the relevant command for your figure-insertion program
% to insert the figure file.
% For example, with the option graphics use
\resizebox{0.5\textwidth}{!}{%
\includegraphics{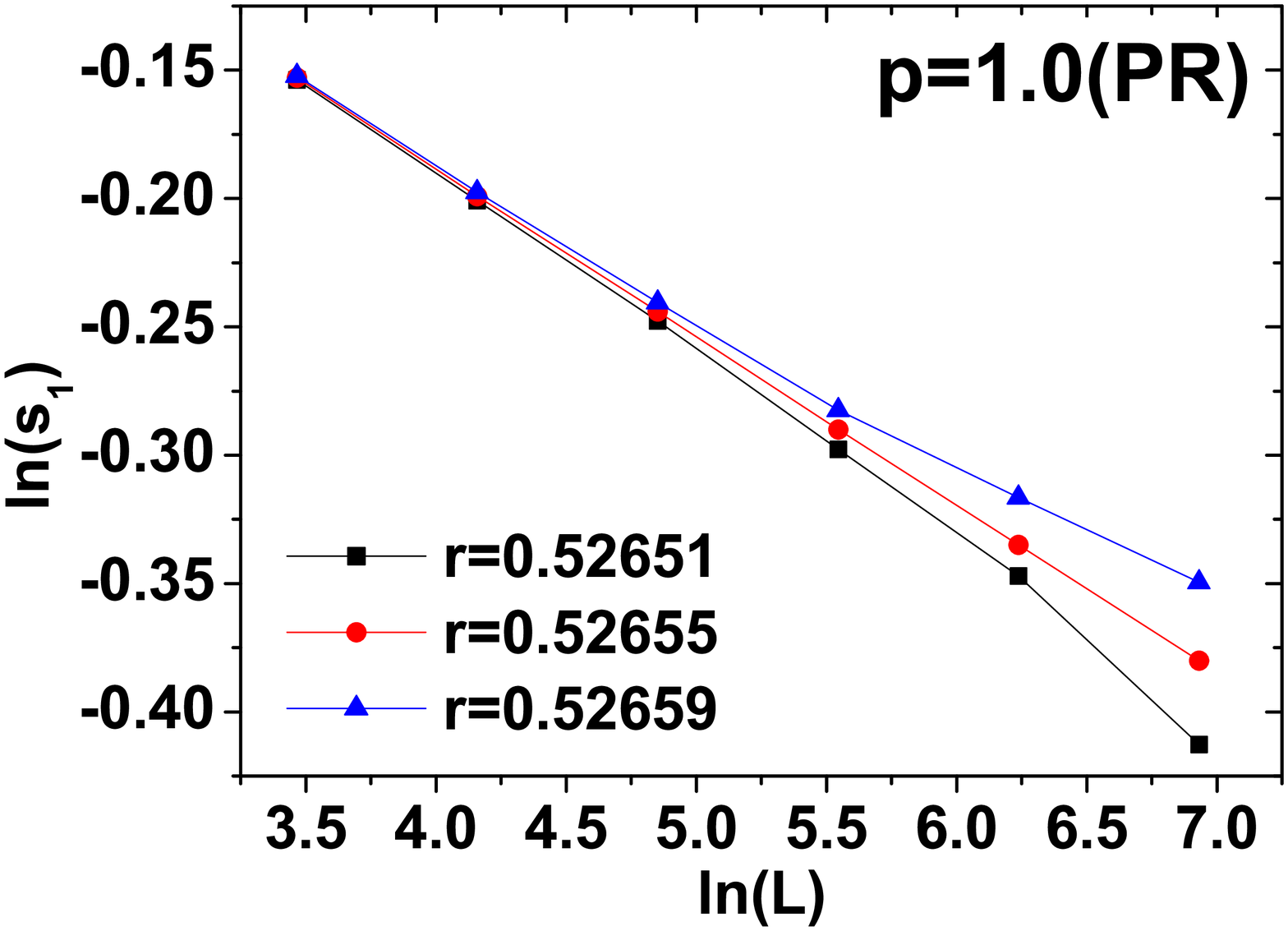}
}
% If not, use
%\vspace{5cm}       % Give the correct figure height in cm
\caption{Log-log plot of the reduced size $s_1$ of the largest  at $p=1.0$
and different $r$. We have chosen linear sizes $L=32,~64,~128,~256,~512$,
and $1024$.}
\label{slopepr}       % Give a unique label
\end{figure}

At $p=1.0$, our model becomes the two-dimensional lattice network
under the Achlioptas process. The ratio $S_2/S_1$ of this model is shown in Fig.\ref{ratiopr} for different system
size $L$. There is a cross-point between $r=0.5265$ and $r=0.5267$, which corresponds to the critical point of this system. Therefore its critical point is at $r_c^{(1)}=0.5266 \pm 0.0001$. The curves of $\ln s_1 (r,L;p)$ at $r=0.52651,0.52655$, and $0.52659$ are shown in Fig.\ref{slopepr}. Their curvatures change with $r$ from negative to positive. The function becomes linear with respect $\ln L$ around $r=0.52655$. So we get the critical reduced edge number $r_c^{(2)}=0.52655 \pm 0.00004$, which agrees with $r_c^{(1)}$ given above. From the slope of $\ln s_1 (r_c,L;p)$ with respect to $\ln L$, the critical exponent ratio $\beta/\nu=0.064$ is obtained. Our results of the critical point and the critical exponent ratio are
in full agreement with the results of Refs.\cite{ZiffPRL09,ChoPRL09,RadicchiPRL09}.
%---------------------------------------------------------------

%---------------------------------------------------------------

\begin{figure}
\resizebox{0.5\textwidth}{!}{%
  \includegraphics{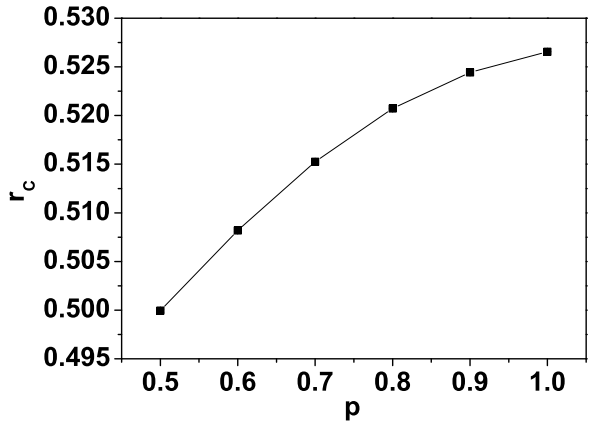}
} \caption{Dependence of the critical reduced edge number $r_c$ on the probability parameter $p$. The error bars of Monte Carlo simulation data are smaller than the symbol.}
\label{rc}
\end{figure}

In Fig.\ref{rc}, we summarize the critical reduced edge numbers $r_c$ of two-dimensional lattice networks under the GAP with different $p$.  It is found that $r_c$ increases with $p$.  At a larger $p$, the continuous percolation phase transition appears at a larger critical reduced edge number $r_c$ and the formation of a giant component is delayed.

\begin{figure}
\resizebox{0.5\textwidth}{!}{%
  \includegraphics{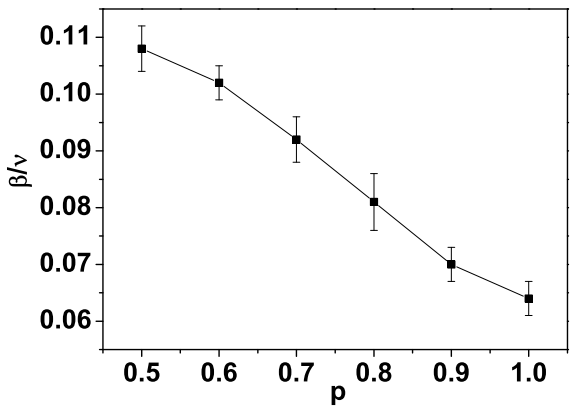}
} \caption{Dependence of the critical exponent ratio $\beta/\nu$ on the probability parameter $p$.}
\label{beta}
\end{figure}

In Fig.\ref{beta}, the dependence of the critical exponent ratio $\beta/\nu$ on $p$ is shown. With the increase of $p$, the ratio $\beta/\nu$ decreases. We will show in the next section that $1/\nu$ increases with $p$. So it can be concluded that the critical exponent $\beta$ decreases with $p$. Smaller $\beta$ indicates the stronger emergence of a giant component in the networks after the percolation transition. Therefore, the increase of $p$ results in the delayed appearance of a continuous percolation phase transition and the formation of a larger giant component at the same time.

\section{Finite-size scaling functions of $S_2/S_1$}
\label{finite-size scaling}

In the last section, we have mentioned that the size ratio $S_2/S_1$ follows the finite-size scaling form in Eq.(\ref{eq:S2/S1}) when $r$ is near the critical point $r_c$. For a given probability parameter $p$, the different curves of $S_2/S_1$ at different $L$ collapse into a finite-size scaling function after using the scaling variable $tL^{1/\nu}$, where $\nu$ is the critical exponent of correlation length. In the following, we will investigate the finite-size scaling function of $S_2/S_1$ for different $p$.

\begin{figure}
\resizebox{0.5\textwidth}{!}{%
\includegraphics{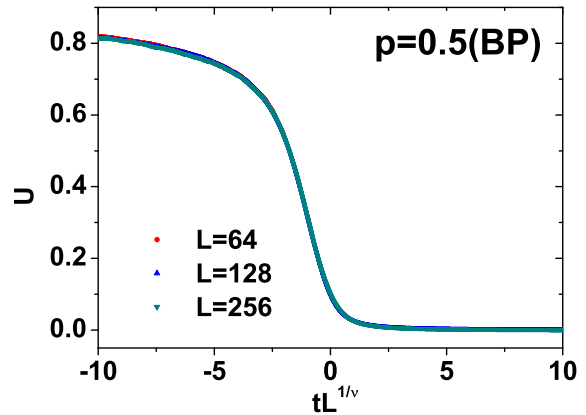} }
\caption{
Finite-size scaling function $U(tL^{1/\nu};p)$ of the ratio $S_2/S_1$ at $p=0.5$. The critical exponent of two-dimensional bond percolation $\nu=4/3$ is taken for the scaling variable.}
\label{scalingbp}
\end{figure}

At $p=0.5$, we use the critical exponent of two-dimensional bond percolation model $\nu=4/3$ \cite{essam} for the finite-size scaling function of $S_2/S_1$. After defining the scaling variable $tL^{1/\nu}$ with this value of $\nu$, our Monte Carlo simulation results at $L=64,~126,~256$ collapse and we get the finite-size scaling function of $S_2/S_1$, which is shown in Fig.\ref{scalingbp}.

\begin{figure}
\resizebox{0.5\textwidth}{!}{%
\includegraphics{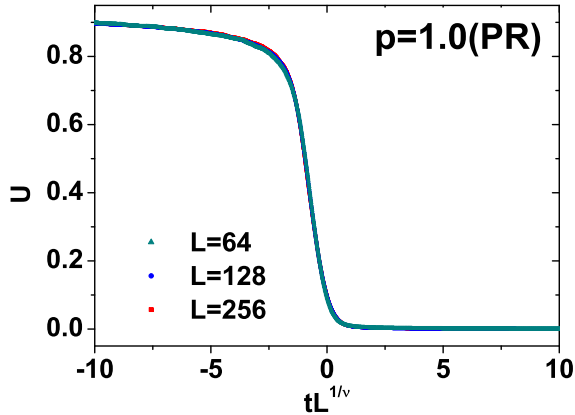} }
\caption{Finite-size scaling function $U(tL^{1/\nu};p)$ of the ratio $S_2/S_1$ at $p=1.0$. The inverse of the critical exponent $1/\nu=0.93$ is taken for the scaling variable.}
\label{scalingpr}
\end{figure}

At $p=1.0$, the critical exponent $\nu$ is unknown. According to the finite-size scaling form in Eq.(\ref{eq:S2/S1}), the curves of $S_2/S_1$ at different system sizes can collapse only when the correct critical exponent $\nu$ is used for the scaling variable. This property can be used also for determining the critical exponent $\nu$. At $1/\nu=0.93$, the curves of $S_2/S_1$ at $L=64,~128,~256$ collapse into its finite-size scaling function, which is shown in Fig.\ref{scalingpr}.

\begin{figure}
\resizebox{0.5\textwidth}{!}{%
  \includegraphics{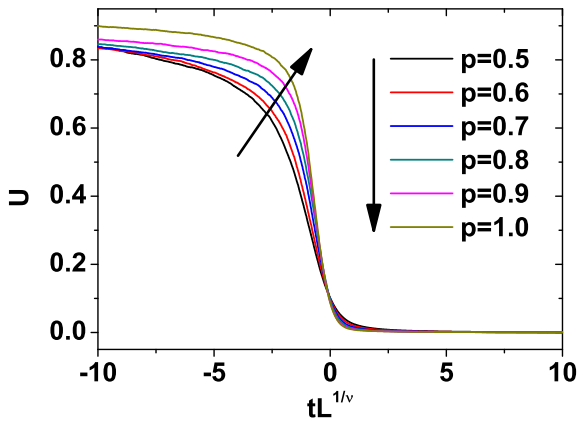}
} \caption{Finite-size scaling functions $U(tL^{1/\nu};p)$ of the ratio $S_2/S_1$ at
$p=0.5,~0.6,~0.7,~0.8,~0.9$, and $1.0$. The corresponding $1/\nu$ for the scaling variable at different $p$ are presented in Fig.\ref{nu} and Tabel \ref{tab1}.}
\label{scalingall}
\end{figure}

\begin{figure}
\resizebox{0.5\textwidth}{!}{%
  \includegraphics{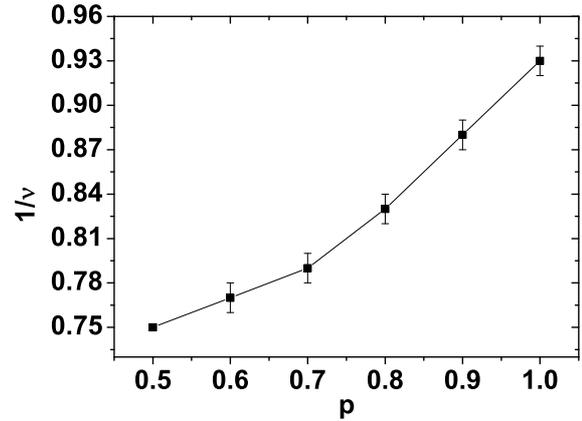}
} \caption{$p$-dependence of the inverse of critical exponent $1/\nu$. The Monte Carlo data at $ p=0.5,~0.6,~0.7,~0.8,~0.9$, and $1.0$ are shown.}
\label{nu}
\end{figure}

In Fig.\ref{scalingall}, we demonstrate the variation of the finite-size scaling function of $S_2/S_1$ with the probability parameter $p$. In the region before the percolation phase transition, the finite-size scaling function of $S_2/S_1$ increases with $p$. The second largest cluster in this region is more important at larger $p$. In the region after the percolation phase transition, the finite-size scaling function of $S_2/S_1$ decreases with $p$. The largest cluster in this region is more dominant at larger $p$. To get the finite-size scaling functions at different $p$, the corresponding exponents of correlation length are determined and presented in Fig.\ref{nu} and Tabel \ref{tab1}.


\begin{figure}
\resizebox{0.5\textwidth}{!}{%
  \includegraphics{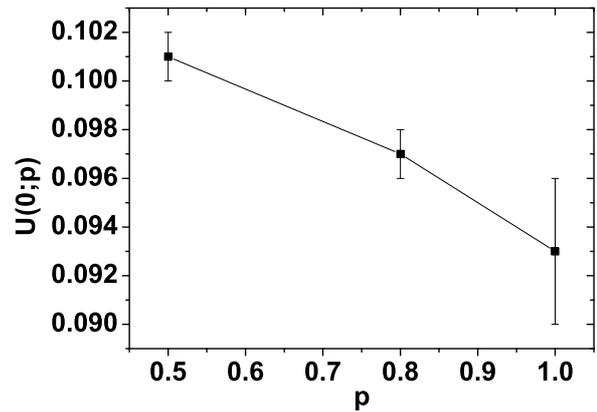}
} \caption{$p$-dependence of the finite-size scaling function $U(tL^{1/\nu};p)$ of the ratio $S_2/S_1$ at the critical point $t=0$.}
\label{U}
\end{figure}

\begin{table}
\caption{Critical reduced edge number $r_c$, ration of critical exponent $\beta/\nu$ and
inverse of critical exponent $1/\nu$ at different probability parameters. We obtain the
critical reduced edge number $r_c^{(1)}$ from $S_2/S_1$ and $r_c^{(2)}$ from
$\ln s_1$.}
\label{tab1}       % Give a unique label
% For LaTeX tables use
\begin{tabular}{lllllll}
\hline\noalign{\smallskip}
%first & second & third  \\
%\noalign{\smallskip}\hline\noalign{\smallskip}
%number & number & number \\
%number & number & number \\
 p  & $r_c^{(1)}$ & $r_c^{(2)}$ & $\beta/\nu$  & $1/\nu$  \\ \hline
$ 0.5$ & 0.5000(2) & 0.5000(4) & 0.108(4) & 0.75 \\
$ 0.6 $ & 0.5082(2) & 0.5082(4) & 0.102(3)& 0.77(1) \\
$ 0.7 $ & 0.5153(3) & 0.5153(2) & 0.092(4) & 0.79(1) \\
$0.8$& 0.5208(1) & 0.5207(2) & 0.081(5) & 0.83(1)\\
$0.9 $& 0.5244(1)& 0.5244(1)& 0.070(3) & 0.88(1)\\
$1.0$& 0.5266(1) & 0.52655(4)& 0.064(3) &  0.93(1)\\
\noalign{\smallskip}\hline
\end{tabular}
% Or use
\end{table}

\section{Universality classes}
\label{universality}
The concept of universality plays a fundamental role in statistical and elementary particle physics \cite{fisher98,zinn-justin}. The universality is characterized by the dimensionality $d$ of the system and by the number $n$ of the components of the order parameter \cite{privman91}. Within a certain $(d,n)$ universality class, the critical exponents are independent of microscopic details and are universal. In a finite-size system near its critical point, there is also universality. For example, the Binder cumulant ratio of magnetization at the critical point is universal. The ratio $S_2/S_1$ here is similar to the Binder cumulant ratio of magnetization. We could suppose that the ratio $S_2/S_1$ at the critical point is also universal and does not depend on microscopic details.

In our previous investigations of the two-dimensional lattice networks under a generalized Achlioptas process, the dimensionality $d$ of systems is fixed and the macroscopic property of their order parameter is unchanged. But we have found in Figs.\ref{beta},\ref{nu} and \ref{U} that the critical exponents $\beta$, $\nu$ and the ratio $S_2/S_1$ at $r_c$ depend on the probability parameter $p$. So the universality of percolation phase transition in these networks is characterized in addition by the probability parameter $p$ of GAP. A different probability parameter $p$ of GAP generates a different probability distribution of configuration. So the probability parameter $p$ is actually related to the macroscopic character of network. For a general classification of universality class in complex networks, further investigations are needed.

\section{Conclusions}

We have investigated the percolation phase transitions in two-dimensional lattice network under a generalized Achlioptas process. In this GAP, we choose randomly two unoccupied edges in a two-dimensional lattice and the edge that minimizes the product of the two connecting cluster sizes is taken with a probability $p$. Our model becomes the two-dimensional bond percolation model at $p=0.5$ and the two-dimensional lattice network under the minority product rule at $p=1$.

The size $S_1$ of the largest cluster in the lattice increases with the edge number $N_r$. When the reduced edge number $r=N_r/N$ is larger than a certain value $r_c$, $S_1$ becomes comparable with the lattice size $N=L^2$. At $r_c$, a giant component emerges and there is a percolation phase transition. From the finite-size scaling analysis of $S_1$ and the ratio $S_2/S_1$, we can conclude that this percolation phase transition is continuous at probability parameter $0.5\le p \le 1$. The critical exponent ratio $\beta/\nu$ can be determined from the power-law behavior of $S_1$ at $r_c$. To obtain the finite-size scaling function of the ratio $S_2/S_1$ from the Monte carlo simulation data of different $L$ with the scaling variable $tL^{1/\nu}$ , the critical exponent of correlation length $\nu$ can be fixed. We find that the critical reduced edge number $r_c$ increases with the probability parameter $p$, which is shown in Fig.\ref{rc}. The critical exponent ratio $\beta/\nu$ and the critical exponent $\nu$ decrease with the probability parameter $p$, as demonstrated in Fig.\ref{beta} and \ref{nu}. Under the GAP with $0.5 < p \le 1$, the formation of larger cluster is suppressed and this suppression increases with $p$. So the formation of a giant component should be delayed at a larger probability parameter $p$. This delay is accompanied then by the stronger emergence of the giant component, which is characterized by smaller $\beta$. It is plausible that $r_c$ increases and $\beta$ decreases with $p$. The finite-size scaling functions of the ratio $S_2/S_1$ are given for different $p$ in Fig.\ref{scalingall}.

Within a certain universality class characterized by the dimensionality of the system and by the number of components of the order parameter, the universal quantities (critical exponents, amplitude ratios, and scaling functions) of different systems are identical. For the two-dimensional lattice networks under a GAP we discuss here, the critical exponents $\beta$, $\nu$ and the ratio $S_2/S_1$ at the critical point depend on the the probability parameter $p$, which has been pointed out above. So the universality class of the percolation phase transition in this model should be characterized in addition by the probability parameter $p$. To understand the universality class of the critical phenomena in networks in general, further investigations are needed. For random networks, we introduce also a generalized Achlioptas process and the phase transitions in these systems are investigated \cite{Fan11}.

This work is supported by the National Natural Science Foundation
of China under grant 10835005.


\begin{thebibliography}{}

\bibitem{StaufferBook}
D.~Stauffer and A.~Aharony, {\em Introduction to Percolation Theory} (Taylor \& Francis, London, 1994).

\bibitem{Solomon} S. Solomon, G. Weisbuch, L. de Arcangelis, N.
Jan, and D. Stauffer, Physica A {\bf 277}, 239 (2000).


\bibitem{Dorogovtsev} S. N. Dorogovtsev, A. V. Goltsev, and J.
F. F. Mendes, Rev. Mod. Phys. {\bf 80}, 1275 (2008).

\bibitem{Achlioptas} D. Achlioptas, R. M. D'Souza, and J. Spencer,
Science {\bf 323}, 1453 (2009).

\bibitem{ZiffPRL09}
R.~M.~Ziff, Phys. Rev. Lett. {\bf 103}, 045701 (2009).

\bibitem{ChoPRL09} Y.~S.~Cho, J.~S.~Kim, J.~Park, B.~Kahng, and D.~Kim, Phys. Rev. Lett. {\bf 103}, 135702 (2009).

\bibitem{RadicchiPRL09} F.~Radicchi and S.~Fortunato, Phys. Rev. Lett. {\bf 103}, 168701 (2009).

\bibitem{ZiffPRE10}
R.~M.~Ziff, Phys. Rev. E {\bf 82}, 051105 (2010).

\bibitem{CostaPRL10}
R.~A.~da~Costa, S.~N.~Dorogovtsev, A.~V.~Goltsev, and J.~F.~F.~Mendes,  Phys. Rev. Lett. {\bf 105}, 255701 (2010).

\bibitem{RiordanArXiv}
O.~Riordan and L.~Warnke, Science {\bf 333}, 322 (2011).

\bibitem{TianArXiv}
L.~Tian and A.~N.~Shi, {\em arXiv:1010.5900} (2010).

\bibitem{GrassbergerArXiv}
P.~Grassberger, C.~Christensen, G.~Bizhani, S.~W.~Son, and M.~Paczuski, {\em arXiv:1103.3728v2}.

\bibitem{RadicchiPRE10}
F.~Radicchi and S.~Fortunato, Phys. Rev. E {\bf 81}, 036110 (2010).

\bibitem{FortunatoArXiv}
S.~Fortunato and F.~Radicchi, {\em arXiv:1101.3567v1} (2011).

\bibitem{newmann1} M. E. J. Newmann and R. M. Ziff, Phys. Rev.
Lett. {\bf 85}, 4104 (2000).

\bibitem{newmann2} M. E. J. Newmann and R. M. Ziff, Phys. Rev.
E {\bf 64}, 016706 (2001).

\bibitem{PF1984}  V. Privman and M. E. Fisher,
Phys. Rev. B  {\bf 30}, 322 (1984).

\bibitem{privman}  V. Privman, {\it Finite Size Scaling and Numerical Simulation of
Statistical Systems}, (World Scientific, Singapore, 1990).

\bibitem{essam} J. W. Essam,{\em Phase Transitions and Critical Phenomena}, edited by C. Domb and J. L. bebowitz (Academie Press, London, 1972), Vol. 2,p. 197.


\bibitem{fisher98} M.~E. Fisher, Rev. Mod. Phys. {\bf 46}, 597 (1974);{\bf 70}, 653 (1998).

\bibitem{zinn-justin} J. Zinn-Justin, {\it Quantum Field Theory and Critical Phenomena}, (Clarendon Press, Oxford,1996).

\bibitem{privman91} V. Privman, A. Aharony, and P.~C. Hohenberg, in {\it Phase Transitions and Critical Phenomena}, edited by C. Domb and J.~L. Lebowitz (Academic, New York, 1991), Vol. 14, p.1.

\bibitem{Fan11} Jingfang Fan, Maoxin Liu, Liangsheng Li, and Xiaosong Chen, to be published.

\end{thebibliography}
\end{document}